\documentclass[draft,jgrga]{AGUTeX}

\usepackage[dvips]{graphicx}
\setkeys{Gin}{draft=false}

\usepackage{multirow}
\authorrunninghead{LI ET AL.}

\titlerunninghead{POST-SHOCK PROPERTIES DEDUCED FROM GEOMAGNETIC INDICES}

\authoraddr{Corresponding author: H. Li,
State Key Laboratory of Space Weather, Center for Space Science and Applied Research, Chinese Academy of Sciences, P.O. Box 8701, Beijing 100190, China. (hli@spaceweather.ac.cn)}

\begin{document}

%% ------------------------------------------------------------------------ %%
%
%  TITLE
%
%% ------------------------------------------------------------------------ %%

\title{Properties of post-shock solar wind deduced from geomagnetic indices responses after sudden impulses}
%
% e.g., \title{Terrestrial ring current:
% Origin, formation, and decay $\alpha\beta\Gamma\Delta$}
%
%% ------------------------------------------------------------------------ %%
%
%  AUTHORS AND AFFILIATIONS
%
%% ------------------------------------------------------------------------ %%

%Use \author{\altaffilmark{}} and \altaffiltext{}

% \altaffilmark will produce footnote;
% matching \altaffiltext will appear at bottom of page.

% \authors{A. B. Smith,\altaffilmark{1}
% Eric Brown,\altaffilmark{1,2} Rick Williams,\altaffilmark{3}
% John B. McDougall\altaffilmark{4}, and S. Visconti\altaffilmark{5}}

%\altaffiltext{1}{Department of Hydrology and Water Resources,
%University of Arizona, Tucson, Arizona, USA.}

\authors{H. Li \altaffilmark{1} and C. Wang \altaffilmark{1}}

\altaffiltext{1}{State Key Laboratory of Space Weather, Center for Space Science and Applied Research, Chinese Academy of Sciences, Beijing, China.}

%% ------------------------------------------------------------------------ %%
%
%  ABSTRACT
%
%% ------------------------------------------------------------------------ %%

% >> Do NOT include any \begin...\end commands within
% >> the body of the abstract.

\begin{abstract}
  Interplanetary (IP) shock plays a key role in causing the global dynamic changes of the geospace environment. For the perspective of Solar-Terrestrial relationship, it will be of great importance to estimate the properties of post-shock solar wind simply and accurately. Motivated by this, we performed a statistical analysis of IP shocks during 1998-2008, focusing on the significantly different responses of two well-used geomagnetic indices (SYMH and AL) to the passive of two types of IP shocks. For the IP shocks with northward IMF (91 cases), the SYMH index keeps on the high level after the sudden impulses (SI) for a long time. Meanwhile, the change of AL index is relative small, with an mean value of only -29 nT. However, for the IP shocks with southward IMF (92 cases), the SYMH index suddenly decreases at a certain rate after SI, and the change of AL index is much significant, of -316 nT. Furthermore, the change rate of SYMH index after SI is found to be linearly correlated with the post-shock reconnection E-field (E$_{KL}$). Based on these facts, an inversion model of post-shock IMF orientation and E$_{KL}$ is developed. The model validity is also confirmed by studying 68 IP shocks in the period of 2009-2013. The inversion accuracy of IMF orientation is 88.24\%, and the inversion efficiency of E$_{KL}$ is as high as 78\%.
\end{abstract}

%% ------------------------------------------------------------------------ %%
%
%  BEGIN ARTICLE
%
%% ------------------------------------------------------------------------ %%

% The body of the article must start with a \begin{article} command
%
% \end{article} must follow the references section, before the figures
%  and tables.

\begin{article}

%% ------------------------------------------------------------------------ %%
%
%  TEXT
%
%% ------------------------------------------------------------------------ %%

\section{Introduction}

As one of the major interplanetary (IP) disturbances, IP shocks link events occurred at the Sun with moderate to extraordinary disturbances at the Earth, and greatly affect the near-Earth environment. Several geoeffectivenesses of IP shocks can be summarized as follows: (1) IP shocks are accompanied by large changes of solar wind dynamic pressure and thus significantly compress the Earth's magnetosphere, causing the sudden impulses (SIs) observed by ground-based magnetometers \citep{Chao and Lepping 1974,Araki 1994}; (2) IP shocks with southward interplanetary magnetic field (IMF) are important interplanetary causes of geomagnetic storms \citep[e.g.][]{Tsurutani et al. 1988,Tsurutani et al. 1992,Gonzalez et al. 1999,Echer et al. 2008a,Echer et al. 2008b}; (3) IP shocks can accelerate the energetic particles during their propagation \citep[e.g.][]{Rao et al. 1967,Lee 1984,Desai et al. 2003} and trigger substorms \citep[e.g.][]{Tsurutani and Zhou 2001,Yamauchi et al. 2006,Yue et al. 2010}. Besides, most of the IP shocks observed within 1 AU can be identified with Interplanetary Coronal Ejections (ICMEs) \citep{Sheeley et al. 1985,Gopalswamy 2006}. Thus, IP shocks can also signal the impending of arrivals of ICMEs.

When an IP shock collides with the Earth's magnetopause, the great compression of magnetosphere will cause a SI of the geomagnetic H-component at middle and low latitudes on the ground. When a SI is followed by a magnetic storm, it is then called storm sudden commencement (SSC). Observations show that almost all SIs are caused by IP shocks \citep{Chao and Lepping 1974,Smith et al. 1986}. The SI amplitude is found to be linearly correlated to the the change in the square root of the solar wind dynamic pressure in many quantitative studies \citep[e.g.][]{Russell et al. 1992,Russell et al. 1994a,Russell et al. 1994b,Wang et al. 2009}. And the rise time of SI generally ranges from 2 to 10 min centered around 4 min \citep{Maeda et al 1962} and is affected by IP shock strength and orientation \citep{Takeuchi et al. 2002,Wang et al. 2006}.

The primary cause of intense geomagnetic storms ($<$ -100 nT) is associated with interplanetary structures with intense and long-duration southward IMF \citep{Gonzalez et al. 1994}. Except for ICME or Magnetic Cloud (MC), another major origin of these structures is the sheath region (SH) behind an IP shock, caused by draping effects or by amplification of the pre-existing southward IMF by shock compression \citep{Gosling et al. 1991}. Thus, IP shocks can have an important role in causing geomagnetic storms. \citet{Echer et al. 2008b} identified the interplanetary causes of intense geomagnetic storms during solar cycle 23 (1996-2006). Based on their storm list, we obtained some simple statistics of intense storms for different drivers (Corotating Interaction Region (CIR) only; ICME only; SH only; combination of SH and ICME, SH+ICME) as shown in Table~1. The intensity of purely SH-driven storm (-172 nT) is significantly greater than that of CIR-driven storm (-118 nT), and is on the same level of purely ICME-driven storm (-166 nT). The storms caused by the combination of SH and ICME have the largest intensity (-223 nT). About 45.2\% of the intense storms are caused by the contribution of IP shocks (SH, SH+ICME). For great storms ($<$ -200 nT), this percentage increases to 72.2\%. Similar conclusion is also made by \citet{Huttunen and Koskinen 2004}.

Based on the SI amplitude observed on the ground, \citet{Wang et al. 2010} made the first effort to estimate IP shock characteristics (such as the change in the square root of solar wind dynamic pressure and the shock speed along the Sun-Earth line) and associated geosynchronous magnetic field variation. Although the estimations are not unique solutions, they still are of particular use in studying events before the space era and could provide proxies for future if we do not have monitoring spacecraft at the L1 point.

The dynamic pressure change across an IP shock affects the large-scale change of magnetosphere structure. Compared with the estimation of post-shock dynamic pressure, the estimations of post-shock IMF orientation and reconnection E-field (E$_{KL}$, proposed by \citep{Kan and Lee 1979}) are of greater importance in magnetospheric storms/substorms. The post-shock IMF orientation determines whether the sheath region can drive a magnetospheric storm/substorm or not, while E$_{KL}$ controls the magnitude of storm/substorm and the duration of substorm growth phase \citep{Li et al. 2010,Li et al. 2013}. Motivated by this, we will develop an inversion model to estimate the properties of post-shock solar wind, such as IMF orientation and E$_{KL}$. This paper is organized as follows: The different responses of two geomagnetic indices after SI for IP shocks with northward and southward IMF are shown in section 2, both from the perspective of case study and of superposed epoch analysis. The inversion model for post-shock solar wind properties is summarized in section 3. The validation of the inversion model is checked up in section 4. At last, section 5 gives the summary.

\section{Different Responses of SYMH/AL Index after SI}

\subsection{Two Comparative Case Study}
Figure 1 shows two IP shocks with northward IMF (left column) and with southward IMF (right column), and the corresponding responses of SYMH/AL index after SIs. The top four panels show successively the sudden changes of IMF magnitude ($B_T$), solar wind bulk speed ($V_{SW}$), solar wind number density ($N_{SW}$), and solar wind temperature ($T$) observed by ACE satellite at L1 point, which represent the occurrence of an IP fast forward shock. The next two panels give the Z component of IMF ($B_Z$) and the reconnection E-field (E$_{KL}$). The bottom two panels show the responses of SYMH index and AL index. Note that, both the SYMH index and AL index have been time-shifted to the shock occurring onset marked as the vertical dashed line.

For the IP shock with northward IMF on June 4 2000, IMF $B_Z$ increased from 4 nT to 10 nT across the shock. The sheath region kept northward IMF for a long time. Meanwhile, E$_{KL}$ had a very minor enhancement and kept on an extreme low level of about 0.2 mV/m. This shock then caused a SI on the ground. The SYMH index increased from -6 nT to 8 nT, and then kept almost invariant. The AL index had no obvious change after the shock collided the magnetosphere. Except for a compression effect of magnetosphere, this IP shock and its sheath region with northward IMF did not caused any magnetic storm or magnetospheric substorm.

For the IP shock with southward IMF on October 28 2001, IMF $B_Z$ decreased from -5 nT to -15 nT across the shock. The sheath region kept southward IMF for more than half an hour. Meanwhile, E$_{KL}$ had a significant enhancement and kept on a high level of about 9 mV/m. Similarly, this shock then caused a SI on the ground, with the SYMH index increasing from -16 nT to 19 nT. Compared to the previous shock event with northward IMF, this shock then led to some different geoeffectivenesses after the SI. The SYMH index decreased suddenly after the peak value of SI at an almost constant rate, which implied the formation of the partial ring current during a magnetic storm as studied by \citet{Li et al. 2011}. The AL index also decreased suddenly and significantly from -200 nT to -700 nT, implying the enhancement of westward electrojet on the high-latitude ionosphere and the possible occurrence of a substorm.

\subsection{Superposed Epoch Analysis}
To check up whether the different responses of SYMH/AL index after SI shown in Figure 1 are common phenomena, a superposed epoch analysis of IP shocks with northward IMF and with southward IMF is performed. During the period of 1998-2008, 91 IP shocks with northward IMF and 92 IP shocks with southward IMF are studied. Note that, the IP shocks with the absolute value of post-shock $B_Z$ less than 2 nT are all excluded to show the differences more clear.

Figure 2 shows the results of superposed epoch analysis of IP shocks with northward IMF (left column) and with southward IMF (right column). Instead of the SYMH index and AL index, the relative values ($\delta$SYMH and $\delta$AL) are processed here, which have subtracted the values at SI rising onset. Except for the bottom two panels, the other panels are the same as those in Figure 1. It is clear that the general changes of $B_T$, $V_{SW}$, $N_{SW}$ and the absolute value of $B_Z$ across the shocks are quite similar for the IP shocks with northward IMF and with southward IMF. However, as shown in Figure 1, the changes of E$_{KL}$ across the shocks and the responses of $\delta$SYMH and $\delta$AL after SIs are obviously different for these two types of IP shocks. For the IP shocks with northward IMF: 1) E$_{KL}$ changed from 0.4 mV/m to 0.7 mV/m across the shocks; 2) the SYMH index kept almost invariant after SIs; 3) the AL index decreased very few after SIs, mostly greater than -100 nT. For the IP shocks with southward IMF: 1) E$_{KL}$ changed significantly from 1.6 mV/m to 4.2 mV/m across the shocks; 2) the SYMH index decreased at a certain rate after SIs; 3) the AL index decreased significantly after SIs.

\section{Inversion Model of Post-shock Solar Wind Properties}
\subsection{Post-shock IMF Orientation Estimation}
As shown in Figure 2, the responses of AL index after SIs are quite different for the IP shocks with northward/southward IMF. To give more details, Figure 3 shows the histogram distribution of $\Delta$AL for these two types of IP shocks. $\Delta$AL = AL$_M$ - AL$_O$, represents the change of AL index after SIs, where AL$_M$ is the mean value of AL index after SI, and AL$_O$ is the AL index at the SI onset. The vertical axis represents the occurrence rate. For the IP shocks with northward IMF, $\Delta$AL ranges mainly from -80 nT to 40 nT, peaking at 0$\sim$40 nT. For the IP shocks with southward IMF, $\Delta$AL ranges mainly from -600 nT to -40 nT, peaking at -600$\sim$-200 nT. The mean value of $\Delta$AL is -29 nT (-316 nT) for the IP shocks with northward (southward) IMF. Inspired by this fact, we try to use $\Delta$AL to be the criterion for estimating post-shock IMF orientation. For practice, the criterion of $\Delta$AL is chosen to be -80 nT. Thus, the inversion model of post-shock IMF orientation is given as follows:
%\begin{linenomath*}
\begin{equation}
 \left\{
\begin{array}{lr}
\Delta \textnormal{AL} \geq \textnormal{-80 nT}, \quad & \textnormal{Northward IMF} \\
\Delta \textnormal{AL} < \textnormal{-80 nT}, \quad & \textnormal{Southward IMF} \\
\end{array}
\right.
\end{equation}
%\end{linenomath*}

Based on this criterion, the post-shock IMF orientations for the 183 IP shocks during 1998-2008 studied in subsection 2.2 is estimated by using the corresponding observations of $\Delta$AL after SIs. The inversion result is shown in Table 2. For 80 out of 91 IP shocks with northward IMF, the post-shock IMF orientations were estimated correctly, with the inversion accuracy of 87.91\%; for 75 out of 92 IP shocks with southward IMF, the post-shock IMF orientations were estimated correctly, with the inversion accuracy of 81.52\%. For all the 183 IP shocks, the inversion accuracy is 84.70\%.

\subsection{Post-shock Reconnection E-field Estimation}
As shown in Figure 2, the SYMH index kept on the high level after SIs for IP shocks with small post-shock reconnection E-field, and decreased at a certain rate for the IP shocks with large post-shock reconnection E-field. Figure 4 shows the relationship between the change rate of SYMH index after SIs ($\delta$SYMH/$\delta$t) and the post-shock reconnection E-filed (E$_{KL}$). To obtain the change rate of SYMH index, we chose the 30-min duration time series of SYMH index after SI and used the least square method to do the linear fitting. The slope of linear fitting is regarded as the change rate of SYMH index. It is clear that $\delta$SYMH/$\delta$t is negatively correlated to the post-shock E$_{KL}$, with the correlation coefficient of -0.86. From this linear relationship, the post-shock E$_{KL}$ can be estimated from $\delta$SYMH/$\delta$t observation. The empirical formula is summarized as follows:
%\begin{linenomath*}
\begin{equation}
\textnormal{E}_{KL}=\left\{
\begin{array}{lr}
0.22-7.70\ \delta \textnormal{SYMH}/\delta \textnormal{t}, \quad & \delta \textnormal{SYMH}/\delta \textnormal{t} \leq 0.028 \\
0.0, \quad & \delta \textnormal{SYMH}/\delta \textnormal{t} > 0.028 \\
\end{array}
\right.
\end{equation}
%\end{linenomath*}

\section{Validation of Inversion Model}
To check up the validity of our inversion model of post-shock solar wind properties, we performed a statistical test of 30 IP shocks with northward IMF and 38 IP shocks with southward IMF identified during the period of 2009-2013.

Table 3 shows the inversion accuracy of post-shock IMF orientation. For 28 out of 30 IP shocks with northward IMF, the post-shock IMF orientations were estimated correctly, with the inversion accuracy of 93.33\%; for 32 out of 38 IP shocks with southward IMF, the post-shock IMF orientations were estimated correctly, with the inversion accuracy of 84.21\%. For all the 68 IP shocks, the inversion accuracy is 88.24\%. On the whole, the post-shock IMF orientation estimation is pretty good.

Figure 5 shows the results of post-shock E$_{KL}$ estimation. The horizontal axis shows the observed E$_{KL}$ and the vertical axis shows the estimated E$_{KL}$. The dashed diagonal line indicates the perfect inversion result. The closer to the diagonal line, the better inversion is. We adopt the inversion efficiency (PE) \citep{Agterberg 1984} to quantify the accuracy of the estimated value compared with the measurement, which is defined as:
%\begin{linenomath*}
\begin{equation}
\textnormal{PE}= 1- \frac{\textnormal{variance of the residual}}{\textnormal{variance of the data}}
\end{equation}
%\end{linenomath*}
The residual is the difference between the measured data and the estimation. The closer of PE to 1, the better inversion is. Although there are some deviations, the estimated E$_{KL}$ are scatted around the diagonal line. The estimations are in good agreement with the observations, with an inversion efficiency of 78\%.

\section{Summary}
As one of the major IP disturbances, IP shock plays a key role in causing the global dynamic changes of the geospace environment. IP shocks are usually accompanied by large changes of solar wind dynamic pressure and thus significantly compress the Earth's magnetosphere. Besides, the sheath regions after IP shocks are another major origins of IP structures with intense and long-duration southward IMF, which are important sources of geomagnetic storms. In addition, IP shocks can also trigger magnetospheric substorms. Thus, it will be of great importance to estimate the properties of post-shock solar wind simply and accurately.

Motivated by this, we performed a statistical analysis of IP shocks with northward (91 cases) and southward (92 cases) interplanetary magnetic field (IMF) during 1998-2008. The responses of two well-used geomagnetic indices, SYMH and AL index, are found to be significantly different to the passive of these two types of IP shocks. For the IP shocks with northward IMF, the SYMH index keeps on the high level after the sudden impulses (SI) for a long time. Meanwhile, the change of AL index is relative small, with the mean value of only -29 nT. However, for the IP shocks with southward IMF, the SYMH index suddenly decreases at a certain rate after SI, and the change of AL index is much significant, of -316 nT. Furthermore, the change rate of SYMH index is found to be linearly correlated with the post-shock reconnection E-field (E$_{KL}$). Based on these facts, an inversion model of post-shock IMF orientation and E$_{KL}$ is developed. For the post-shock IMF orientation, the inversion model is summarized as follows:
%\begin{linenomath*}
\begin{equation}
 \left\{
\begin{array}{lr}
\Delta \textnormal{AL} \geq \textnormal{-80 nT}, \quad & \textnormal{Northward IMF} \\
\Delta \textnormal{AL} < \textnormal{-80 nT}, \quad & \textnormal{Southward IMF} \\
\end{array}
\right.
\end{equation}
%\end{linenomath*}
And for the post-shock reconnection E-field, the inversion model is summarized as follows:
%\begin{linenomath*}
\begin{equation}
\textnormal{E}_{KL}=\left\{
\begin{array}{lr}
0.22-7.70\ \delta \textnormal{SYMH}/\delta \textnormal{t}, \quad & \delta \textnormal{SYMH}/\delta \textnormal{t} \leq 0.028 \\
0.0, \quad & \delta \textnormal{SYMH}/\delta \textnormal{t} > 0.028 \\
\end{array}
\right.
\end{equation}
%\end{linenomath*}
The validity of this inversion model is also verified by studying 68 IP shocks in the period of 2009-2013. The inversion accuracy of IMF orientation is 88.24\%, and the inversion efficiency of E$_{KL}$ is as high as 78\%.

These estimations of post-shock IMF orientation and reconnection E-field are an extension of our previous work \citep{Wang et al. 2010}, which made the first effort to estimate IP shock characteristics, such as the change in the square root of solar wind dynamic pressure, the shock speed along the Sun-Earth line, and associated geosynchronous magnetic field variation, from SI amplitude observed on the ground. These estimations are of particular use in studying events before the space era and could provide proxies for future times if we do not have a solar wind monitoring spacecraft at the L1 point.
%%% End of body of article:

%%%%%%%%%%%%%%%%%%%%%%%%%%%%%%%%
%% Optional Appendix goes here
%
% \appendix resets counters and redefines section heads
% but doesn't print anything.
% After typing  \appendix
%
% \section{Here Is Appendix Title}
% will show
% Appendix A: Here Is Appendix Title
%
%%%%%%%%%%%%%%%%%%%%%%%%%%%%%%%%%%%%%%%%%%%%%%%%%%%%%%%%%%%%%%%%
%
% Optional Glossary or Notation section, goes here
%
%%%%%%%%%%%%%%
% Glossary is only allowed in Reviews of Geophysics
% \section*{Glossary}
% \paragraph{Term}
% Term Definition here
%
%%%%%%%%%%%%%%
% Notation -- End each entry with a period.
% \begin{notation}
% Term & definition.\\
% Second term & second definition.\\
% \end{notation}
%%%%%%%%%%%%%%%%%%%%%%%%%%%%%%%%%%%%%%%%%%%%%%%%%%%%%%%%%%%%%%%%
%
%  ACKNOWLEDGMENTS

\begin{acknowledgments}
We acknowledge the use of 64-s ACE data combined the magnetic field and solar wind plasma from Coordinated Data Analysis Web (CDAWeb). We also thank the Kyoto World Data Center for providing the SYMH/AL indices. This work was supported by 973 program 2012CB825602, NNSFC grants 41204118 and 41231067, and in part by the Specialized Research Fund for State Key Laboratories of China.
\end{acknowledgments}

\end{article}

%% Enter Figures and Tables here:

% When submitting articles through the GEMS system:
% COMMENT OUT ANY COMMANDS THAT INCLUDE GRAPHICS.
%
% DO NOT USE \psfrag or \subfigure commands.
%
% Figure captions go below the figure.
% Table titles go above tables; all other caption information
%  should be placed in footnotes below the table.

% DRAFT figure/table, including eps graphics
%
% \begin{figure}
% \noindent\includegraphics[width=20pc]{samplefigure.eps}
% \caption{Caption text here}
% \end{figure}
% \end{document}
%
% \begin{table}
% \caption{}
% \end{table}
%
% ---------------
% TWO-COLUMN figure/table
%
% \begin{figure*}
% \noindent\includegraphics[width=39pc]{samplefigure.eps}
% \caption{Caption text here}
% \end{figure*}
%
% \begin{table*}
% \caption{Caption text here}
% \end{table*}
%
% ---------------
% EXAMPLE TABLE
%
%\begin{table}
%\caption{Time of the Transition Between Phase 1 and Phase 2\tablenotemark{a}}
%\centering
%\begin{tabular}{l c}
%\hline
% Run  & Time (min)  \\
%\hline
%  $l1$  & 260   \\
%  $l2$  & 300   \\
%  $l3$  & 340   \\
%  $h1$  & 270   \\
%  $h2$  & 250   \\
%  $h3$  & 380   \\
%  $r1$  & 370   \\
%  $r2$  & 390   \\
%\hline
%\end{tabular}
%\tablenotetext{a}{Footnote text here.}
%\end{table}

% See below for how to make landscape/sideways figures or tables.
\newpage
\begin{table}
\caption{Statistics of intense magnetic storms ($<$ -100 nT) during solar cycle 23 (1996-2006) for different drivers: (1) CIR only; (2) ICME only; (3) SH only; (4) SH+ICME.}
\centering
\renewcommand{\arraystretch}{1.9}
\setlength{\arrayrulewidth}{0.5pt}
\setlength{\doublerulesep}{0.05pt}
\begin{tabular*}{30pc}{l@{\extracolsep{\fill}}c@{\extracolsep{\fill}}c@{\extracolsep{\fill}}c@{\extracolsep{\fill}}c@{\extracolsep{\fill}}}
\hline
\hline
 & CIR & ICME & SH & SH+ICME \\
\hline
\hline
Case number & 12 & 33 & 22 & 15 \\
Percentage & 14.6\% & 40.2\% & 26.9\% & 18.3\% \\
Storm intensity & -118 nT & -166 nT & -172 nT & -223 nT \\
Great storm\tablenotemark{a} & 0 (0.0\%)\tablenotemark{b} & 5 (27.8\%) & 6 (33.3\%) & 7 (38.9\%) \\
\hline
\hline
\end{tabular*}
\tablenotetext{a}{Great storm represents the magnetic storm with the peak SYMH index $<$ -200 nT.}
\tablenotetext{b}{The value precedes the bracket is the case number, and the value in the bracket represents its percentage.}
\end{table}

\newpage
\begin{table}
\caption{Inversion accuracy of post-shock IMF orientation for the IP shocks during 1998-2008.}
\centering
\renewcommand{\arraystretch}{1.9}
\setlength{\arrayrulewidth}{0.5pt}
\setlength{\doublerulesep}{0.05pt}
\begin{tabular*}{30pc}{l@{\extracolsep{\fill}}c@{\extracolsep{\fill}}c@{\extracolsep{\fill}}}
\hline
\hline
 & Northward (Estimated) & Southward (Estimated)\\
\hline
\hline
Northward (observed) & 80 (87.91\%) & 11 (12.09\%) \\
Southward (observed) & 17 (18.48\%) & 75 (81.52\%) \\
\hline
\hline
\end{tabular*}
\end{table}

\newpage
\begin{table}
\caption{Inversion accuracy of post-shock IMF orientation for the IP shocks during 2009-2013.}
\centering
\renewcommand{\arraystretch}{1.9}
\setlength{\arrayrulewidth}{0.5pt}
\setlength{\doublerulesep}{0.05pt}
\begin{tabular*}{30pc}{l@{\extracolsep{\fill}}c@{\extracolsep{\fill}}c@{\extracolsep{\fill}}}
\hline
\hline
 & Northward (Estimated) & Southward (Estimated)\\
\hline
\hline
Northward (observed) & 28 (93.33\%) & 2 (6.67\%) \\
Southward (observed) & 6 (15.79\%) & 32 (84.21\%) \\
\hline
\hline
\end{tabular*}
\end{table}

\newpage
\begin{figure}
\centering
\noindent\includegraphics[width=18pc]{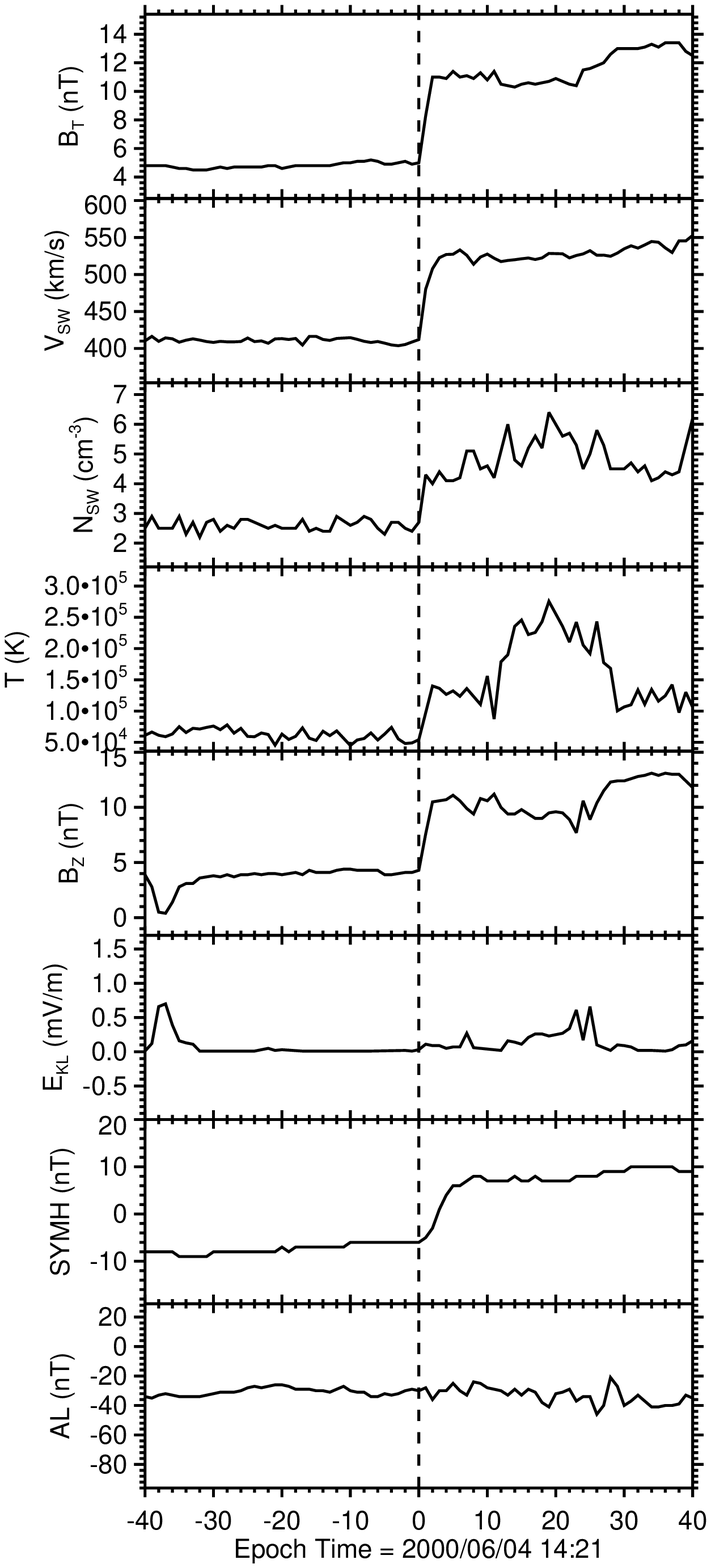}
\noindent\includegraphics[width=18pc]{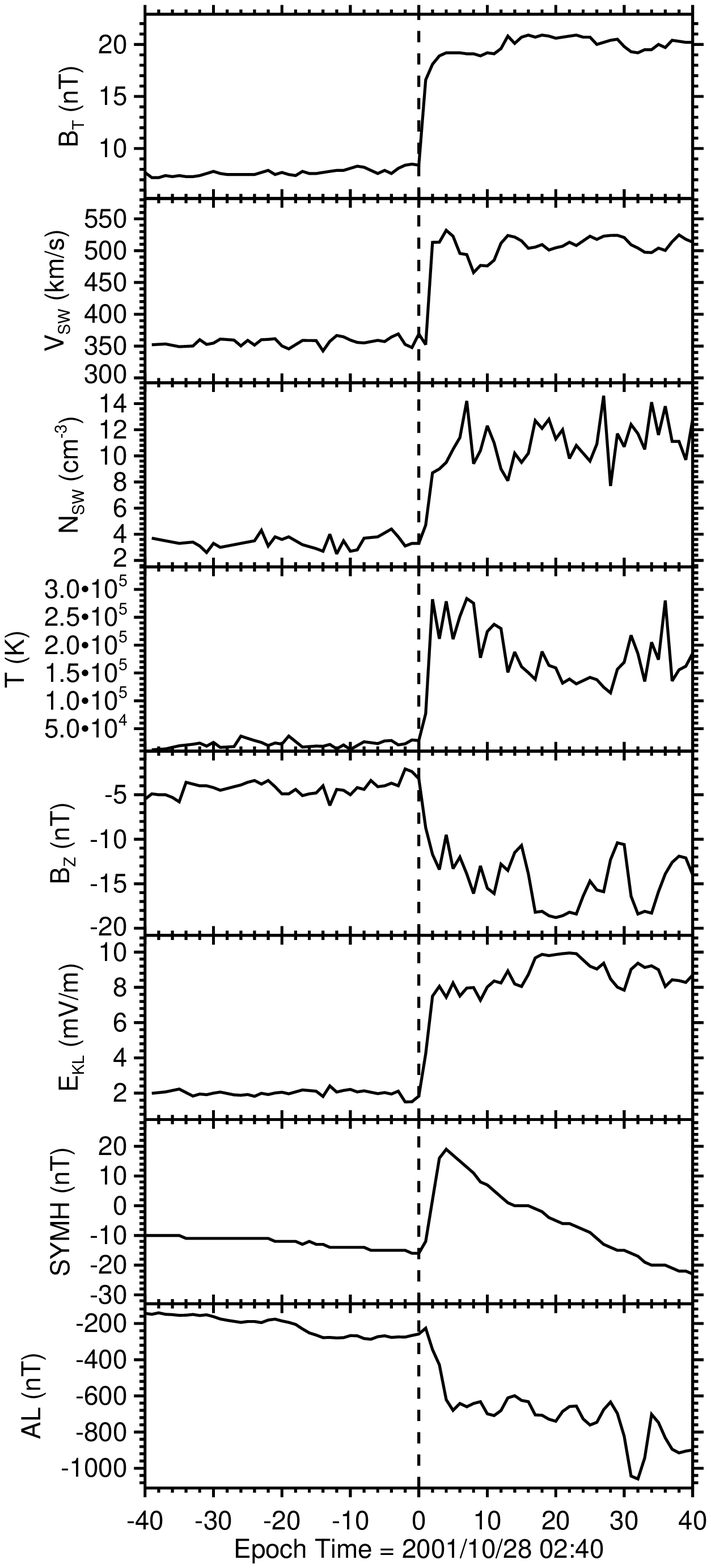}
\caption{Two comparative events of IP shocks with northward IMF on June 4, 2000 (left column) and with southward IMF on October 28, 2001 (right column). Note that, both the SYMH index and AL index have been time-shifted to the shock occurring onset marked as the vertical dashed line.}
\end{figure}

\newpage
\begin{figure}
\centering
\noindent\includegraphics[width=18pc]{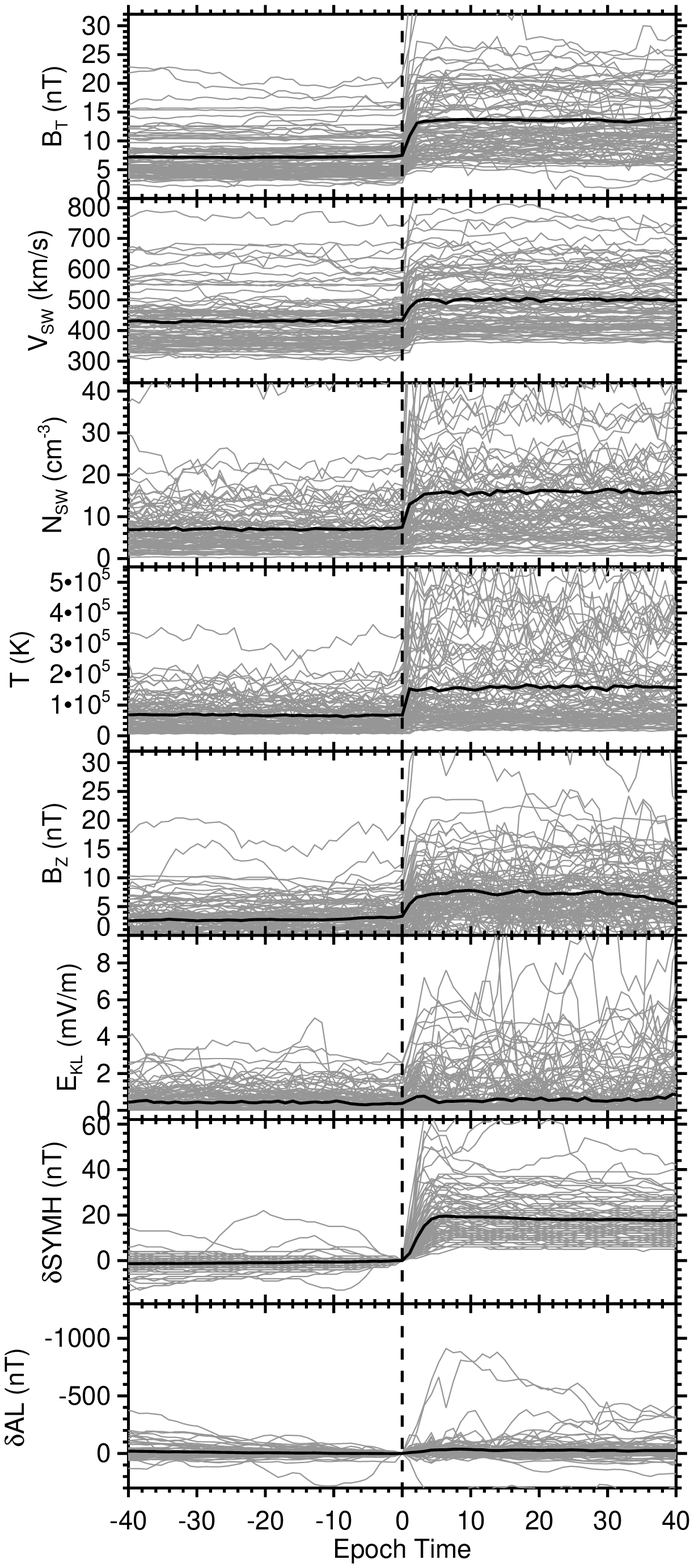}
\noindent\includegraphics[width=18pc]{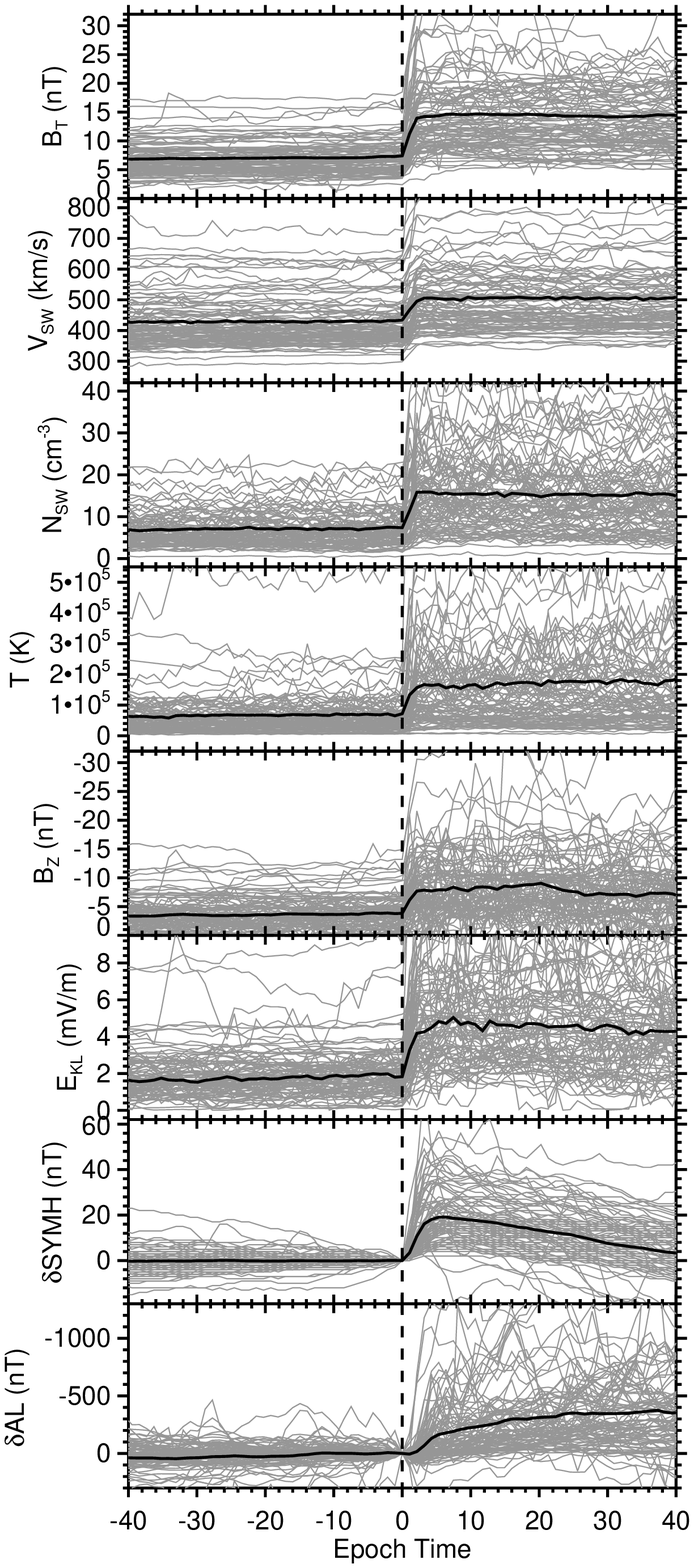}
\caption{Superposed epoch analysis of IP shocks with northward IMF (left column) and with southward IMF (right column). The gray line is the observation for each IP shock, and the black line represent the mean value. Instead of the SYMH index and AL index, the relative values ($\delta$SYMH and $\delta$AL) are processed here, which have subtracted the values at SI rising onset. Note that, both the $\delta$SYMH and $\delta$AL have been time-shifted to the shock occurring onset marked as the vertical dashed line.}
\end{figure}

\newpage
\begin{figure}
\centering
\noindent\includegraphics[width=26pc]{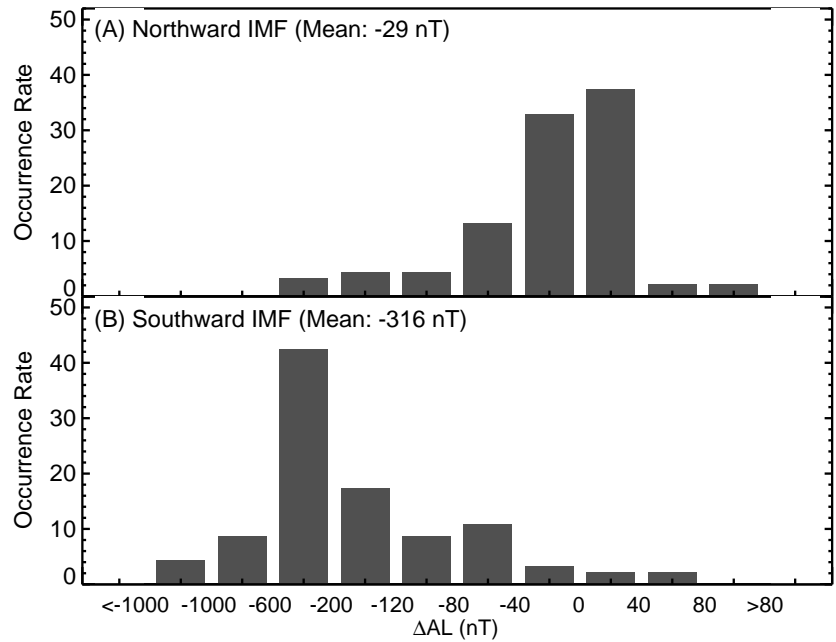}
\caption{Histogram distribution of $\Delta$AL for two types of IP shocks: (A) IP shocks with northward IMF; (B) IP shocks with southward IMF.}
\end{figure}

\newpage
\begin{figure}
\centering
\noindent\includegraphics[width=26pc]{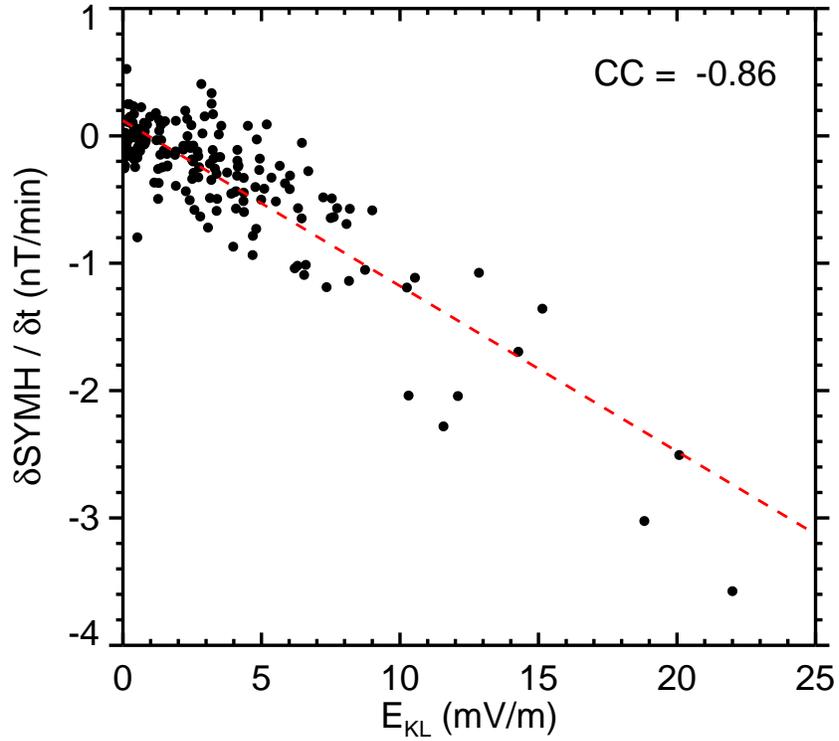}
\caption{Relationship between the change rate of SYMH index after SIs and the post-shock reconnection E-field. The red line represents the linear fitting result. CC represents the correlation coefficient.}
\end{figure}

\newpage
\begin{figure}
\centering
\noindent\includegraphics[width=26pc]{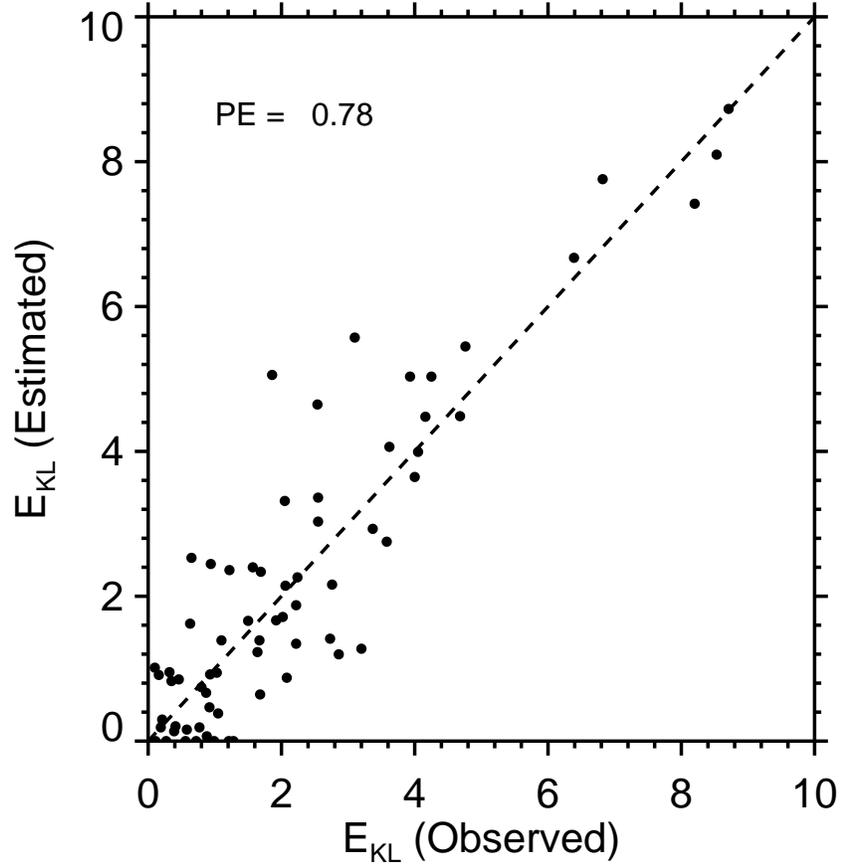}
\caption{Results of post-shock reconnection E-field estimation. The dashed diagonal line indicates the perfect inversion. PE represents the inversion efficiency. }
\end{figure}

\end{document}